\documentclass[aps,prl,reprint,superscriptaddress,showpacs]{revtex4-1}

\usepackage{amsmath,amssymb,graphicx,units,ulem}
\usepackage[usenames,dvipsnames]{color}

\renewcommand{\vec}[1]{\boldsymbol{#1}}

\begin{document}
\title{Mott-insulator-to-superconductor transition in a two-dimensional 
superlattice}

\author{Rubem Mondaini}
\affiliation{Department of Physics, The Pennsylvania State University, 
University Park, PA 16802, USA}

\author{Predrag Nikoli\'{c}}
\affiliation{School of Physics, Astronomy and Computational Sciences,
George Mason University, Fairfax, VA 22030, USA and \\
Institute for Quantum Matter at Johns Hopkins University, 
Baltimore, MD 21218, USA}

\author{Marcos Rigol} 
\affiliation{Department of Physics, The Pennsylvania State University, 
University Park, PA 16802, USA}

\begin{abstract}
We use quantum Monte Carlo and exact diagonalization calculations to 
study the Mott-insulator to superconductor quantum phase transition in 
a two-dimensional fermionic Hubbard model with attractive interactions 
in the presence of a superlattice potential. The model introduced offers
unique possibilities to study such transitions in optical lattice experiments. 
We show that, in regimes with moderate to strong interactions, the transition 
belongs to the $3D$-$XY$ universality class. We also explore the character of 
the lowest energy charge excitations in the insulating and superconducting 
phases and show that they can be fermionic or bosonic depending on the 
parameters chosen.
\end{abstract}
\maketitle

\section{Introduction}
The traditional microscopic approach to understanding superconductivity scrutinizes 
the pairing instability of a ``parent'' normal state \cite{Bardeen1957}. If this normal 
state is not a conventional state of weakly interacting electrons (i.e., a Fermi liquid 
or a band-insulator), then the emerging superconductivity is not ordinary either. Many 
superconductors fall in this category, including organic, heavy-fermion, and all 
high-temperature ones (cuprates and iron-based). Non-trivial electron correlations 
behind superconductivity are most famously seen in the ``pseudogap'' state of cuprates, 
and are very difficult to understand from experimental observations \cite{Ando1995, Corson1999,
Ong2001, Davis2002, Valla2006, Gomes2007, Kohsaka2008, Ghir2012}. This is where ultracold-atom 
systems, which are highly tunable, are expected to help. However, the temperatures
required to realize the $d$-wave pairing of cuprates remain prohibitively low 
for current ultracold atom experiments \cite{esslinger_review_10}. Even obtaining 
long-range antiferromagnetic correlations in a three-dimensional Mott insulator remains 
a challenge \cite{jordens_strohmaier_08,schneider_hackermuller_08,hart_duarte_14}.

With these challenges in mind, it is desirable to design $s$-wave paired states 
that would make the ``pseudogap'' physics accessible to current experiments with 
ultracold fermions. Here, we study a zero 
temperature Mott insulator of bound Cooper pairs, which gives rise to the $s$-wave 
analogue of a pseudogap state, as well as to a superconductor upon
changing lattice parameters. We consider a Fermi-Hubbard model in the square lattice 
in the presence of a superlattice potential, with Hamiltonian
\begin{eqnarray}
  &&\hat H = -t \sum_{\langle {\vec {i\,j}} \rangle ,\sigma}
  ( \hat c^{\dagger}_{\vec{i}\sigma} \hat c^{}_{\vec{j}\sigma}
  + \text{H.c.}) 
  -t'\sum_{\langle\langle {\vec {i\,j}} \rangle\rangle ,\sigma}
  (\hat c_{\vec{i}\sigma}^\dagger \hat c_{\vec{j}\sigma}^{}
  + \text{H.c.}) \nonumber \\
  &&+U \sum_{\vec i} \left(\hat n_{\vec i\uparrow}-\frac{1}{2}\right)
  \left(\hat n_{\vec i \downarrow}-\frac{1}{2}\right)
  +\Delta \sum_{\vec{i},\sigma}(-1)^{\left(i_x + i_y\right)}
  \hat n_{\vec i\sigma},\quad\
\label{eq:Hamiltonian}
\end{eqnarray}
where $c^{\dagger}_{\vec{i}\sigma}$ ($c^{}_{\vec{i}\sigma}$) are the fermionic 
creation (annihilation) operators at site $\vec {i}$, with (pseudo-)spin 
$\sigma=\uparrow,\downarrow$, and $\hat n_{\vec i\sigma}=
\hat c^{\dagger}_{\vec{i}\sigma} \hat c^{}_{\vec{i}\sigma}$ are the 
corresponding site occupation operators. The nearest and next-nearest hopping 
amplitudes are denoted by $t$ and $t'$ [$\langle {\vec {i\,j}} \rangle$ 
and $\langle\langle {\vec {i\,j}} \rangle\rangle$ indicate sums over 
nearest and next-nearest neighbor sites $\vec {i}$ and $\vec {j}$], respectively,
the strength of the on-site attractive interaction by $U<0$, and 
of the staggered potential by $\Delta$.

In experiments, attractive interactions between atoms can be generated using 
Feshbach resonances \cite{chin_grimm_review_10}, optical superlattices can be 
created using arrays of laser beams~\cite{Sebby-Strabley2006,Foelling2007,Lee2007}, 
and periodically modulated optical lattices allow to control the relative 
amplitudes and phases between nearest and next-nearest neighbor hopping parameters 
\cite{Lignier2007,Zenesini2009}. A recent experimental realization of the Haldane 
model exemplifies these capabilities \cite{Jotzu2014}. We stress as a caveat to 
be kept in mind that, in experiments using Feshbach resonances, the single-band 
description based on Eq.~\eqref{eq:Hamiltonian} fails when the gap between different 
Hubbard bands is not much larger than the other energy scales involved in the 
problem~\cite{Duan2005, Diener2006, Chin2006}. 

To show what makes the model in Eq.~\eqref{eq:Hamiltonian} special to study 
Mott-insulator to superconductor phase transitions, we analyze it in two limits. 
In the non-interacting limit ($U=0$), $\hat{H}$ can be 
diagonalized in \textbf{k}-space, which unveils two bands,
$E(\mathbf{k})=-4t'\cos{k_x}\cos{k_y}\pm\sqrt{\epsilon^2_{\mathbf{k}}+\Delta^2}$,
where the reduced Brillouin zone is given by $|k_x+k_y|\leq\pi$ and $|k_x-k_y|\leq\pi$, 
and $\epsilon_{\mathbf{k}}=-2t\left[\cos(k_x)+\cos(k_y)\right]$ is the dispersion 
relation in the presence only of nearest neighbors hopping. For $t'< t/\sqrt{2}$,
an indirect gap opens for $\Delta_c = 2t'$ [between ($\pm\pi$,0) in the lower 
band and $(\pm\pi/2,\pm\pi/2)$ in the upper band].  For $t'> t/\sqrt{2}$, an 
indirect gap opens for $\Delta_c = 4t'-t^2/t'$ [between ($\pm\pi$,0) in the lower 
band and (0,0) in the upper band]. $\Delta_c$ is the critical value of 
$\Delta$ for the formation of a band insulator at half filling, i.e., 
a finite value of $t'$ stabilizes a metallic state for nonzero 
values of $\Delta$.

The other important limit is the one in which $U/t\neq0$ but $t'=0$. Recalling 
that the attractive Hubbard Hamiltonian can be mapped onto a repulsive one 
by the down-spin particle-hole transformation~\cite{Micnas1990}, 
$\hat c^{}_{\vec{i}\downarrow}\leftrightarrow \left(-1\right)^{i_x + i_y}
\hat c^{\dagger}_{\vec{i}\downarrow}$, the staggered potential transforms as
$\Delta\sum_{\vec i\sigma}(-1)^{\left(i_x + i_y\right)} \hat n_{\vec i\sigma}
\rightarrow h\sum_{\vec i}(-1)^{\left(i_x + i_y\right)} \hat S^z_i$, 
with $\hat S^z_i=\left(\hat n_{\vec i\uparrow}-\hat n_{\vec i\downarrow}\right)/2$
and $h=2\Delta$. Therefore, at half-filling, the staggered potential in the 
attractive model is equivalent to a staggered $z$-magnetic field in the 
repulsive one. For $h=0$, the ground state of the repulsive Hubbard model 
is an $SU(2)$ symmetric Mott insulator that exhibits long-range antiferromagnetic 
correlations. Those translate onto long-range $s$-wave superconducting order and 
charge density-wave order in the attractive model (i.e., a supersolid) \cite{Micnas1990}. 
An infinitesimal $h$ breaks $SU(2)$ symmetry and the ground state of the repulsive model 
becomes an $S^z$ antiferromagnet, which translates onto a (charge density-wave) 
Mott insulator in the attractive case, i.e., superconductivity is destroyed for 
any nonzero value of $\Delta$. For large values of $h$, this insulator can be 
understood to be the result of pinning the pairs to the sites with 
energies $-\Delta$, which precludes transport.

Now if one takes the $U=0$ limit with $t'\neq 0$ and $\Delta<\Delta_c$ 
as the starting point, adding weak attractive interactions generates 
superconductivity, i.e., contrary to the $t'=0$ case, 
superconductivity is possible for $\Delta\neq0$. Increasing $\Delta$, 
one can then destroy the superconductor in favor of a Mott-insulator. Such a 
transition is the focus of this work. It can be driven in real time in 
ultracold fermion experiments by tuning lattice parameters. This is to be 
contrasted to the Mott-insulator to superconductor transition in the cuprates, 
which requires changing doping, i.e., the filling would need to be changed 
in real time to drive such a transition in optical lattices.

We study Hamiltonian \eqref{eq:Hamiltonian} in $L$$\times$$L$ lattices
using two unbiased computational approaches: zero-temperature (projector) 
determinantal quantum Monte Carlo (PDQMC) \cite{Muramatsu1999,Assaad2002} 
and Lanczos exact diagonalization (ED). We focus on half-filled systems 
($n=\langle n_{\vec i\uparrow}\rangle+\langle n_{\vec i\downarrow}\rangle=1.0$,
$\langle n_{\vec i\uparrow}\rangle=\langle n_{\vec i\downarrow}\rangle$), 
except when analyzing the nature of the charge excitations. The projector 
parameter in the PDQMC calculations was set to $\Theta t=40$, ensuring 
that the we obtain ground state properties for lattices with up to 256 sites,
while the imaginary time discretization step was taken to be $\delta\tau=0.1$. 
In the ED calculations, we used translational symmetries, which allowed 
us to solve lattices with up to 16 sites. $t=1$ sets the energy scale in all
results reported in what follows.

\section{Results}
\begin{figure}[!tb] 
  \includegraphics[width=1\columnwidth]{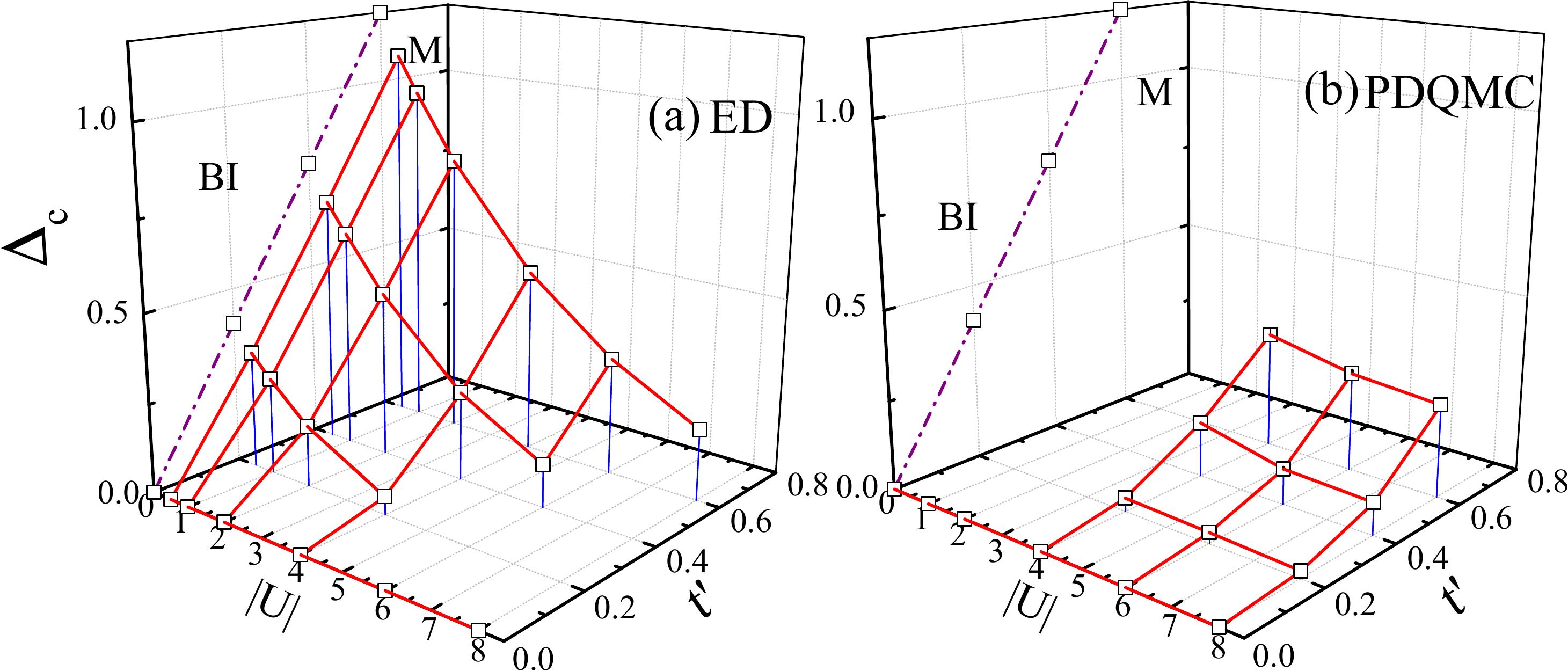}
  \caption{(Color online) Phase diagram of Eq.~\eqref{eq:Hamiltonian} 
  obtained via ED (a) and PDQMC (b). For $U/t=0$, the dashed line marks the 
  boundary between the metallic and band-insulating phases obtained analytically. 
  For nonzero $U$, the surface formed by connecting the points (which report 
  $\Delta_c$) delimits the insulating ($\Delta>\Delta_c$) and superconducting
  ($\Delta<\Delta_c$) phases. Comparing (a) and (b), one can see that $\Delta_c$ 
  is overestimated in the ED calculations due to finite-size effects.}
  \label{fig:phase_diagram}
\end{figure}

Figure~\ref{fig:phase_diagram} depicts the phase diagram for 
Eq.~\eqref{eq:Hamiltonian}, obtained using ED [Fig.~\ref{fig:phase_diagram}(a)] 
and PDQMC [Fig.~\ref{fig:phase_diagram}(b)], as given by the $\Delta_c$ 
necessary to drive the superconductor to Mott-insulator transition 
as a function of $|U|$ and $t'$. Important features visible in 
Fig.~\ref{fig:phase_diagram} are, (i) $\Delta_c$ decreases with 
increasing $|U|$, and, (ii) as expected from the discussion in the 
noninteracting limit, $\Delta_c$ increases with increasing $t'$. 
The first trend can be understood as both the attractive interaction 
and the staggered potential favor local pair formation and, consequently, 
reduce long-range order when $\Delta\neq0$ and $|U|$ is increased.
The second trend follows from the fact that the delocalization 
promoted by $t'$ competes with the pinning induced by $\Delta$ and 
$U$, and enhances superconductivity.

In the ED calculations, in order to determine $\Delta_c$ for the superconductor to 
Mott-insulator transition at fixed $U$ and $t'$, we use the ground-state fidelity metric
\cite{Zanardi2006,Venuti2007,Zanardi2007,Yang2007,rigol_shastry_09,Varney2011,Jia2011}
\begin{equation}
\mathrm{g}(\Delta)=\frac{2}{L^2}\frac{
1-|\langle\Psi_0(\Delta)|\Psi_0(\Delta+\delta\Delta)\rangle|}{\left(\delta\Delta\right)^2},
\label{eq:fidelity_metric}
\end{equation}
where $|\Psi_0(\Delta)\rangle$ is the ground-state wavefunction of the Hamiltonian 
for a given staggered on-site energy $\Delta$ and $\delta\Delta$ is chosen 
to be small enough that the results for $\mathrm{g}(\Delta)$ are 
independent of its value. $\mathrm{g}(\Delta)$ is expected to exhibit a 
diverging (with increasing system size) maximum as one crosses a second order phase transition 
\cite{Zanardi2006,Venuti2007,Zanardi2007,Yang2007,rigol_shastry_09,Varney2011,Jia2011}.

\begin{figure}[!tb] 
  \includegraphics[width=0.98\columnwidth]{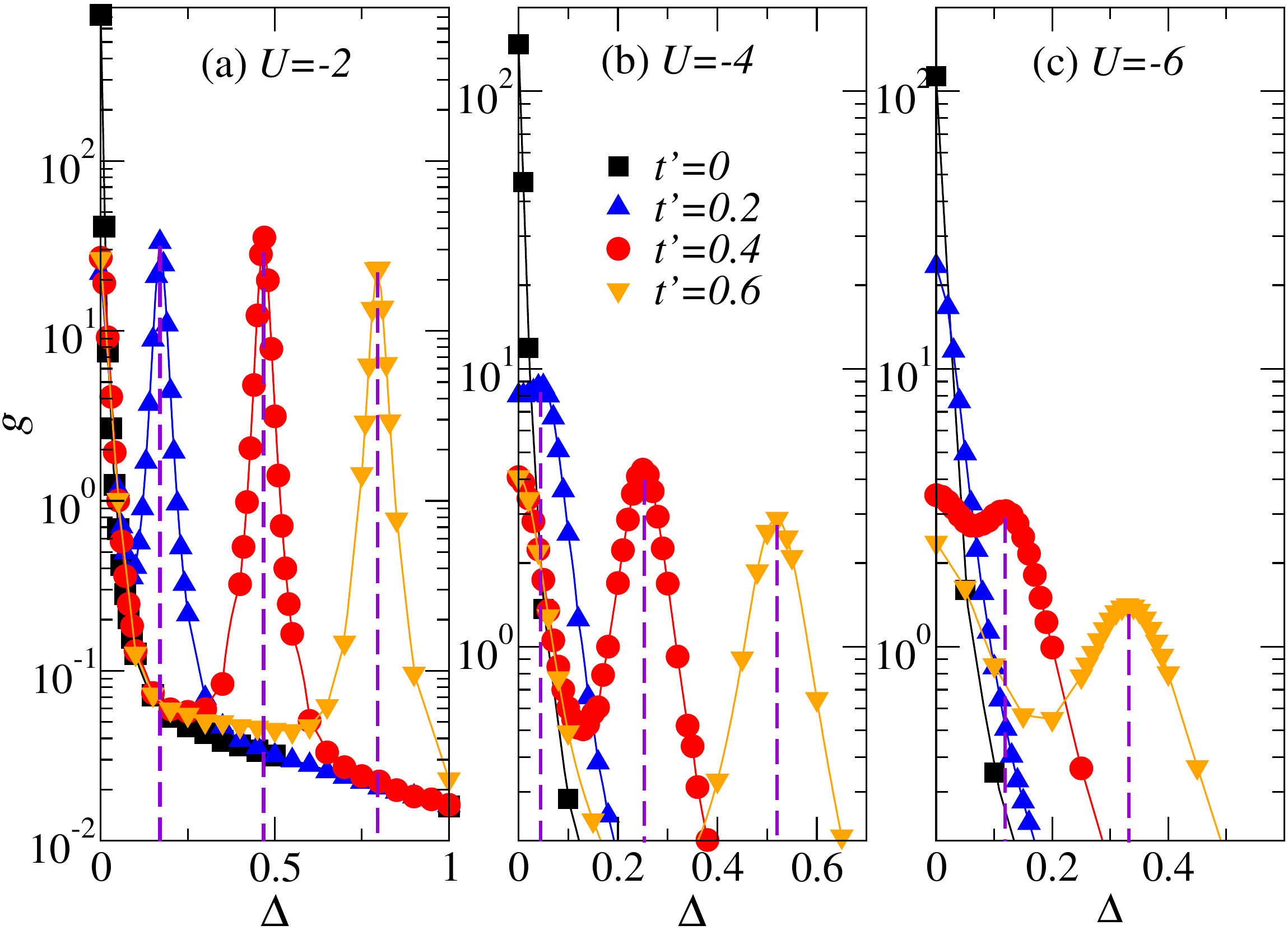}
  \vspace{-0.1cm}
  \caption{(Color online) ED results for the fidelity metric vs $\Delta$ in $4\times4$ 
  lattices at half-filling, for fours values of $t'$ ($t'=0,\,0.2,\,0.4$ and $0.6$), 
  and three values of $U$ [$U=-2$ (a), $-4$ (b) and $-6$ (c)]. For $t'=0$, the only 
  peak seen in $g$ is the one associated with the destruction of superconductivity 
  for any nonzero value of $\Delta$, whereas for finite values of $t'$ a second 
  peak appears at finite values of $\Delta$ signaling the superconductor
  to Mott-insulator transition.}
  \label{fig:fidelity}
\end{figure}

Figures~\ref{fig:fidelity}(a)--\ref{fig:fidelity}(c) depict the fidelity metric for 
different values of the onsite interaction ($U=-2,\,-4$ and $-6$, respectively) 
and for four values of $t'$ ($t'=0.0,\,0.2,\,0.4$ and $0.6$). For $t'=0$ and all 
values of $U$, one can see that there is a single peak in $g$ for $\Delta\simeq0$. 
This peak signals the supersolid to Mott-insulator transition previously discussed
for the limit $U\neq0$ but $t'=0$. The signature of such a transition can still 
be seen for $t'\neq0$ in the form of a peak at $\Delta\simeq0$ with a height that 
decreases with increasing $t'$ (notice the log scale in the $y$-axes). A second 
peak then emerges for $t'\neq0$ signaling the superconductor to Mott-insulator transition
with increasing $\Delta$. We take the position of the maximum of this peak as the 
value of $\Delta_c$ predicted by ED. The compilation of these peak 
positions provides the phase diagram reported in Fig.~\ref{fig:phase_diagram}(a).
Notice that, with increasing $|U|$, the positions of the peaks for a given value of 
$t'$ move towards smaller values of $\Delta$. They also become broader and at some 
point merge with the one at $\Delta\simeq0$. At that point, we cannot determine $\Delta_c$ 
using ED. This is why the phase diagram in Fig.~\ref{fig:phase_diagram}(a) is missing
points for large $|U|$ and small $t'$ values.

\begin{figure}[!tb] 
  \includegraphics[width=0.99\columnwidth]{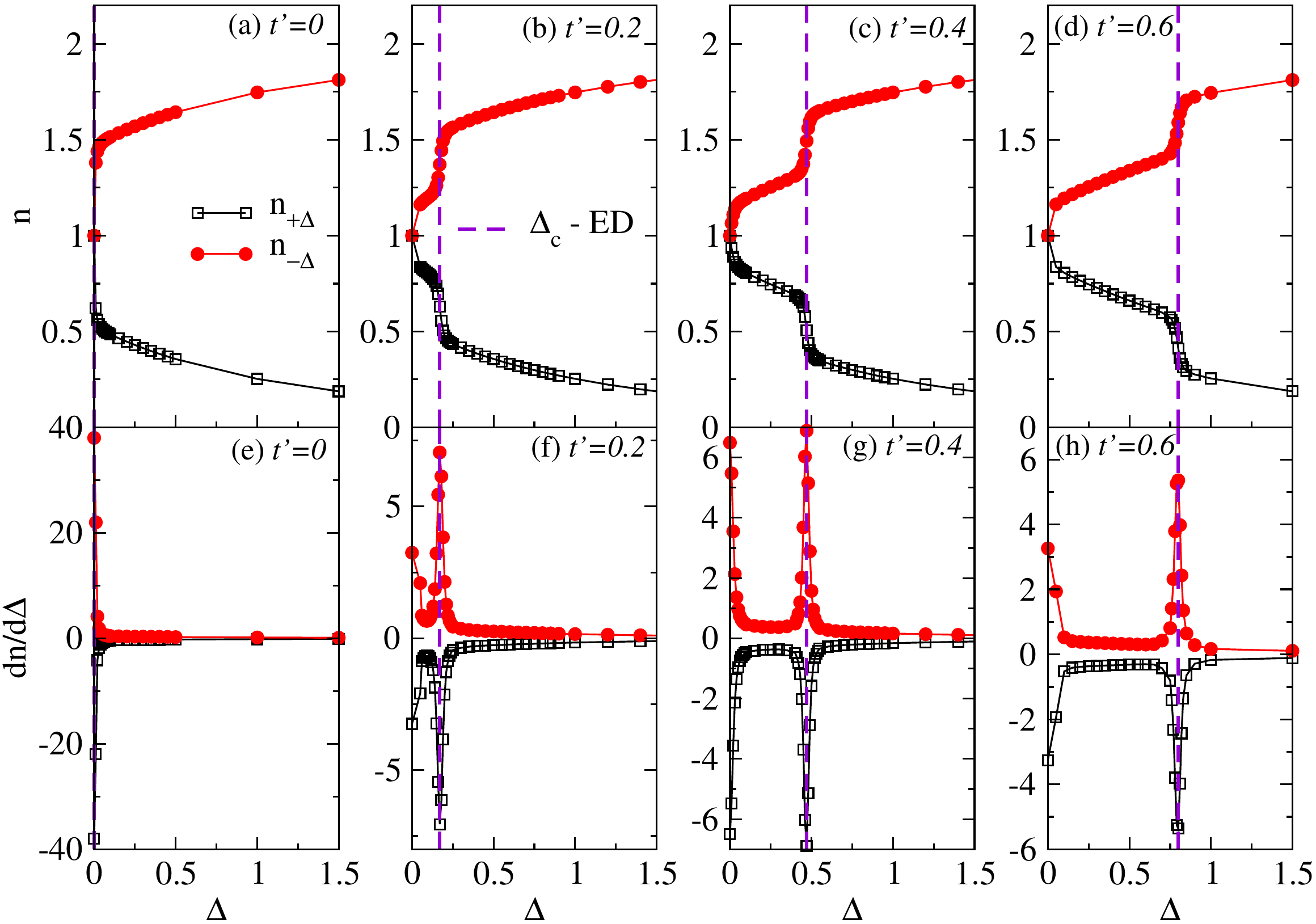}
  \caption{(Color online) (a)--(d) Site occupation 
  $\langle \hat{n}_{\vec{i}\uparrow} + \hat{n}_{\vec{i}\downarrow}\rangle$ 
  in each sublattice (labeled by $n_{+\Delta}$ and $n_{-\Delta}$) as a function 
  of $\Delta$ for different values of $t'$ and $U=-2$. Vertical dashed lines 
  mark the superconductor-insulator transition obtained through the fidelity 
  metric. (e)--(h) Derivative of the site occupations in (a)--(d) 
  with respect to $\Delta$. All results were obtained by means of ED in a 
  $4\times4$ lattice at half-filling.}
  \label{fig:densities_ed}
\end{figure}

Other quantities also show clear signatures of the transition.
In the presence of the superlattice potential, the sublattices 
forming the bipartite square lattice possess different onsite energies and, 
consequently, the site occupation is different in the two site 
species. Figures~\ref{fig:densities_ed}(a)--\ref{fig:densities_ed}(d) display 
the site occupation in each sublattice as a function of $\Delta$ for $U=-2$ 
and several values of $t^\prime$. As expected, as $\Delta$ increases, 
the difference between the site occupation in the sublattices increases.
Remarkably, there is a sharp increase in this difference that occurs exactly
at the value of $\Delta$ for which the fidelity metric predicts the 
superconductor to Mott-insulator transition. This sharp increase leads to
a sharp peak in the derivative of the site occupation with respect to 
$\Delta$, see Figs.~\ref{fig:densities_ed}(e)--\ref{fig:densities_ed}(h), 
which resembles the peak seen in the fidelity metric.

Another observable that exhibits clear signatures of the superconductor to 
Mott-insulator transition is the binding energy
\begin{equation}
E_{b}=2E_0(n+1)-E_0(n+2)-E_0(n),
\label{eq:pair_binding}
\end{equation}
where $E_0(n)$ correspond to the ground state energy of a system with $n$ 
fermions. 

\begin{figure}[!b] 
  \includegraphics[width=0.99\columnwidth]{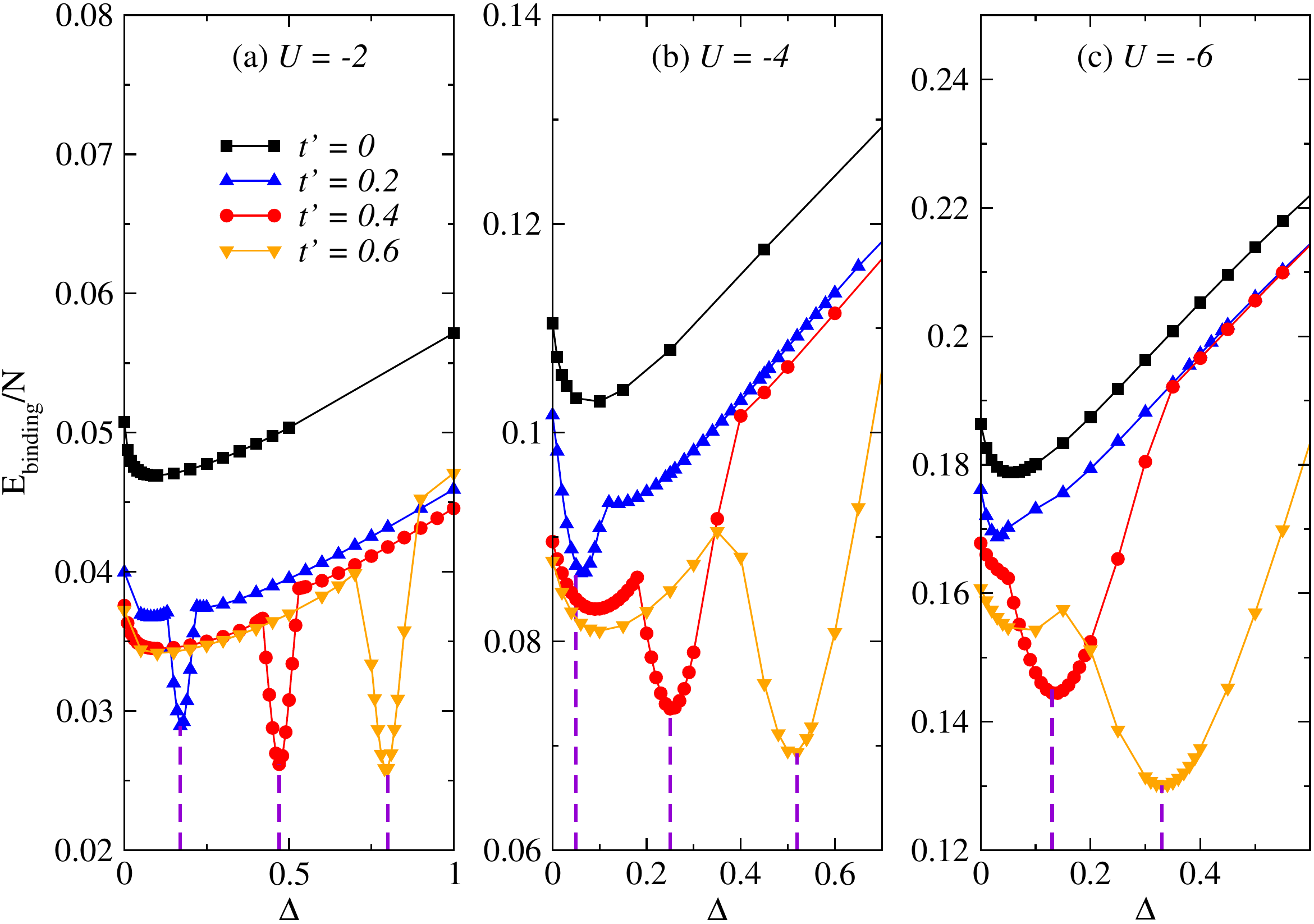}
  \caption{(Color online) Binding energy as a function of $\Delta$ for 
  $U = -2, -4$ and $-6$, and different values of $t'$. The dashed lines, 
  which coincide with the minimum in the dips of the binding energy, 
  report the values of $\Delta_c$ provided by the fidelity metric. 
  All results were obtained by means of ED in a 
  $4\times4$ lattice at half-filling.}
  \label{fig:binding_energy}
\end{figure}

Figure~\ref{fig:binding_energy} shows the binding energy vs 
$\Delta$ for different values of $U$ and $t'$. The trend is similar in all of 
them: for small values of $\Delta$, before the superconductor to Mott-insulator 
transition for nonzero $t'$ takes place, the energy associated with the pairs 
decreases, and then a dip occurs exactly at the transition point as detected by 
the fidelity metric. For larger values of $\Delta$, in the Mott insulating 
phase, the binding energy steadily increases with increasing $\Delta$.

\begin{figure}[!tb] 
  \includegraphics[width=0.99\columnwidth]{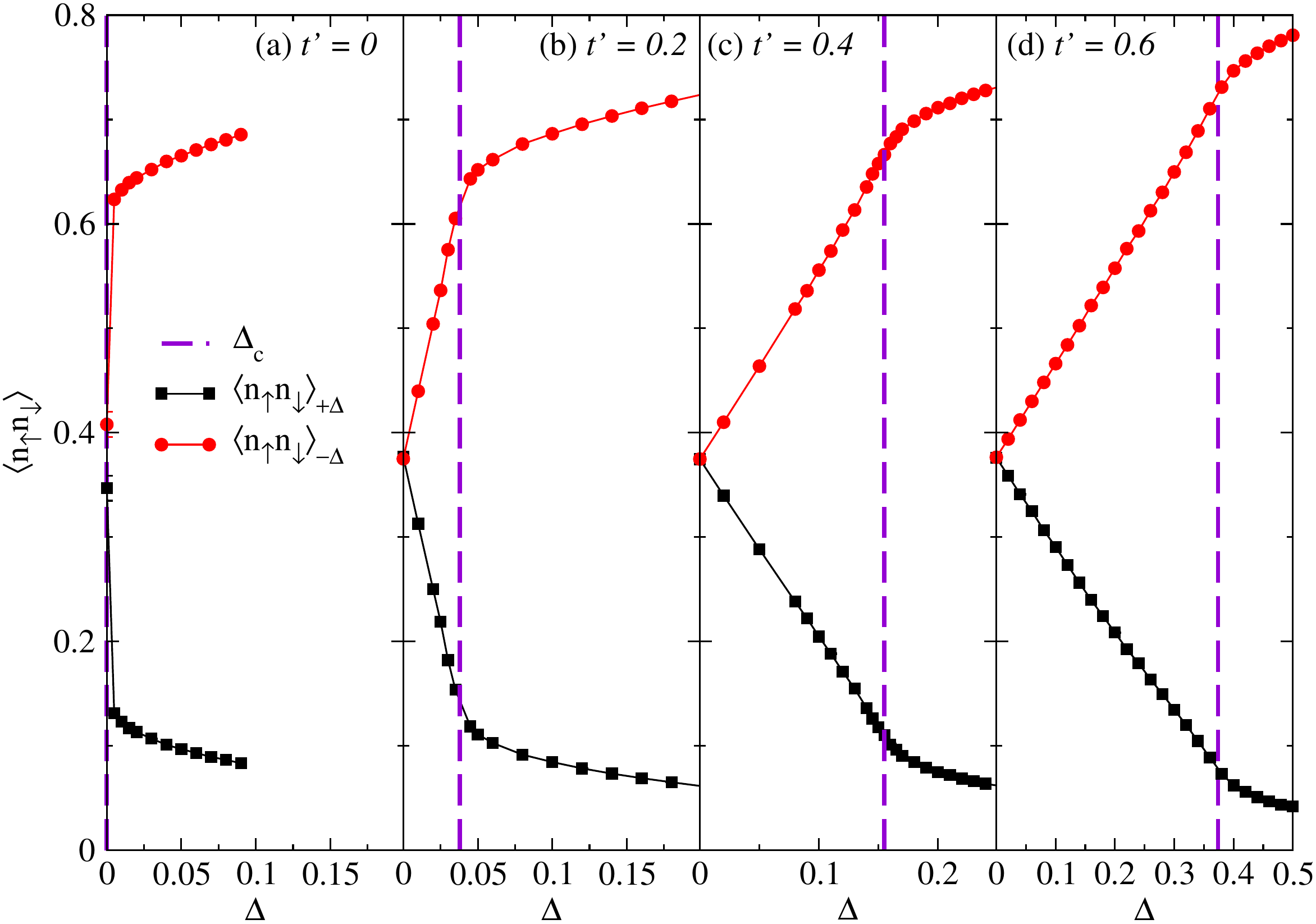}
  \caption{(Color online) Double occupancy in the two sublattices as a function 
of $\Delta$ for $U=-4$ and different values of $t'$. These results were 
obtained using PDQMC in a $14\times14$ lattice at half filling. The vertical 
dashed lines depict the value of $\Delta_c$ as obtained in the finite-size 
scaling analysis of the pair structure factor.}
  \label{fig:docc}
\end{figure}

In turn, similar robustness against the selection of the observable used to 
characterize the transition is seen in the PDQMC calculations of much larger 
lattice sizes than those amenable to exact diagonalization. An observable of 
much interest in experiments with ultracold fermions in optical lattices is the 
double occupancy. It was used, e.g., in Ref.~\cite{jordens08} to detect the 
Mott insulating phase when increasing the onsite repulsion between fermions. 
In Fig.~\ref{fig:docc}, we plot the double occupancy in the two sublattices
vs $\Delta$ for $U=-4$ and different values of $t'$ in a $14\times14$ lattice.
A kink can be seen in the behavior of this observable around (slightly after) 
the critical value of $\Delta$ obtained in the finite size scaling of the pair 
structure factor (see the following subsection).

\begin{figure}[!t] 
  \includegraphics[width=0.98\columnwidth]{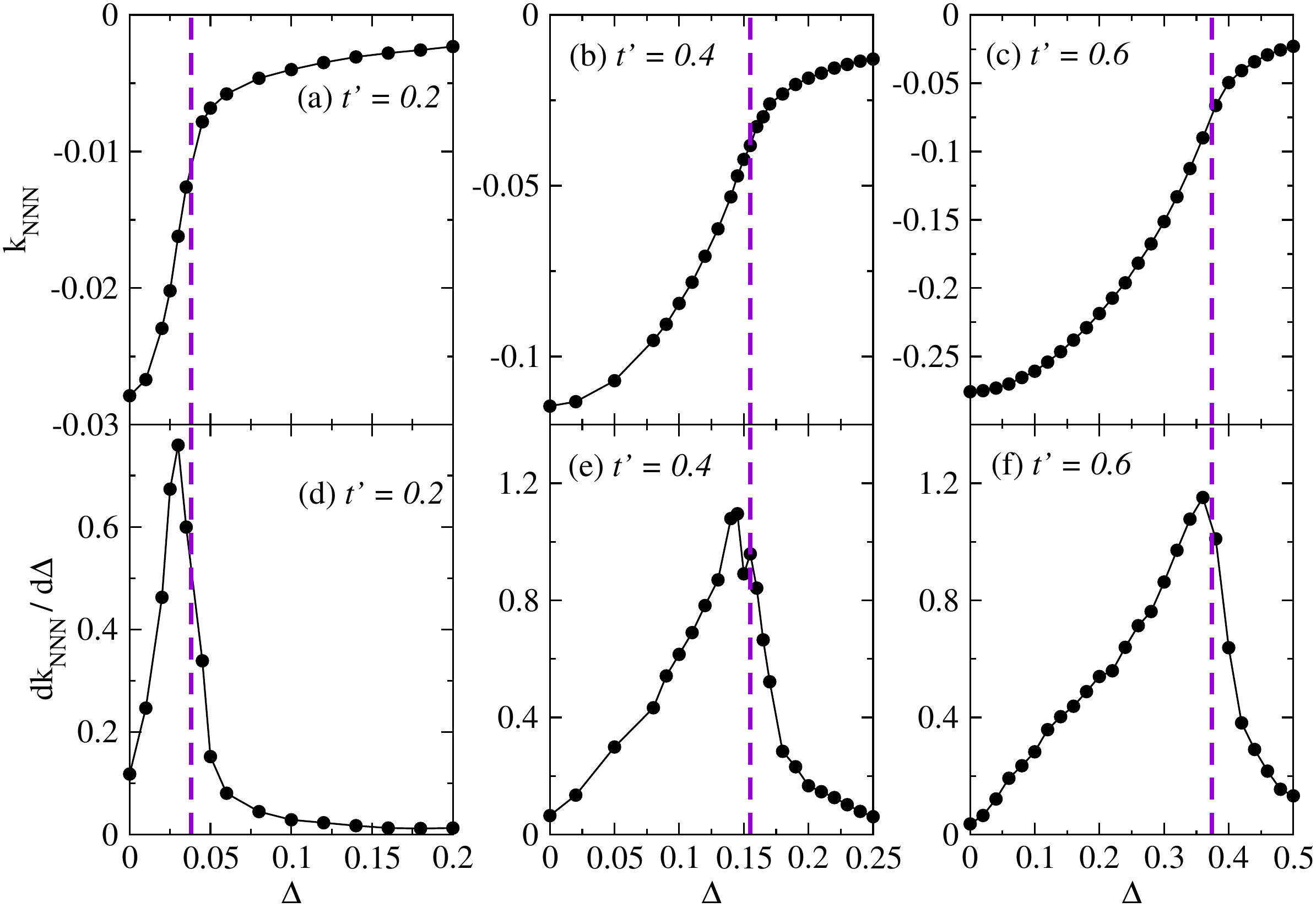}
  \caption{(Color online) The kinetic energy associated with next-nearest
neighbor hopping as a function of $\Delta$ for different values of $t'$ in 
systems with $U=-4$ and $14\times 14$ sites. The vertical dashed lines show 
the value of $\Delta_c$ as obtained in the finite-size scaling analysis of 
the pair structure factor.}
  \label{fig:k_nnn_w_diff}
\end{figure}

Another local quantity that exhibits a clear signature of the superconductor 
to Mott-insulator transition is the kinetic energy associated with next-nearest
neighbor hoppings, i.e., $k_{NNN} = -(t'/ N) \sum_{\langle\langle 
{\vec {i\,j}} \rangle\rangle ,\sigma}\langle \hat c_{\vec{i}\sigma}^\dagger 
\hat c_{\vec{j}\sigma}^{}+\hat c_{\vec{j}\sigma}^\dagger \hat 
c_{\vec{i}\sigma}^{} \rangle$), where $N=L\times L$. As shown in 
Figs.~\ref{fig:k_nnn_w_diff}(a)--\ref{fig:k_nnn_w_diff}(c), the absolute
value of $k_{NNN}$ decreases with increasing $\Delta$. This observable also
exhibits a kink right after $\Delta_c$ as predicted by the scaling 
analysis. The derivatives of $k_{NNN}$ with respect to $\Delta$, shown in 
Figs.~\ref{fig:k_nnn_w_diff}(d)--\ref{fig:k_nnn_w_diff}(f), exhibit clear 
peaks about $\Delta_c$ with maxima right before the value of $\Delta_c$ reported
in the phase diagram.

Summarizing, the ED and the PDQMC results for various observables studied 
indicate the occurrence of the superconductor to Mott-insulator transition at 
approximately the same value of $\Delta$ as the fidelity metric. Hence, the 
results reported in the phase diagram (Fig.~\ref{fig:phase_diagram}) are robust 
against the selection of the observable.

\subsection{Universality class of the transition}

The PDQMC calculations have the advantage that they allow us to study 
much larger lattice sizes and, after a proper finite-size scaling analysis, 
determine $\Delta_c$ in the thermodynamic limit. We take the pair structure 
factor $P=\sum_{i,j}\langle\hat P^{}_j\hat P_i^\dagger\rangle$, with 
$\hat P_i=\hat c^{}_{\vec{i}\uparrow}\hat c^{}_{\vec{i}\downarrow}$, to 
be the order parameter to locate the superconductor to Mott-insulator transition
(as mentioned in the previous section, other observables give similar results). 
The limit $|U|/t\gg1$, for $t'=0$, provides guidance on the nature 
of the superconductor to Mott-insulator transition for strong attractive interactions.
Second-order perturbation theory in $t/|U|$ reveals that Eq.~\eqref{eq:Hamiltonian} 
becomes equivalent to a Hamiltonian for hard-core bosons with site 
creation (annihilation) operator $\hat b^\dagger_i=\hat P^\dagger_i$ 
($\hat b^{}_i=\hat P_i$) \cite{Robaszkiewicz1981,Micnas1990} in a superlattice. 
The phase diagram for the latter model was studied in Refs.~\cite{Hen2009,Hen2010} 
in two (2D) and three (3D) dimensions. At half-filling, this model exhibits 
a superfluid to Mott-insulator transition with increasing $\Delta$ that 
belongs to the ($d$+1)-$XY$ universality class 
\cite{Hen2009,Hen2010}, like the integer filling Mott transition in the Bose-Hubbard 
model \cite{Fisher1989}. The addition of $t'$ to the hard-core boson model does not 
break its particle-hole symmetry. Hence, it does not qualitatively 
change the phase diagram in Refs.~\cite{Hen2009,Hen2010}.

\begin{figure}[!tb] 
  \centering
  \includegraphics[width=0.99\columnwidth]{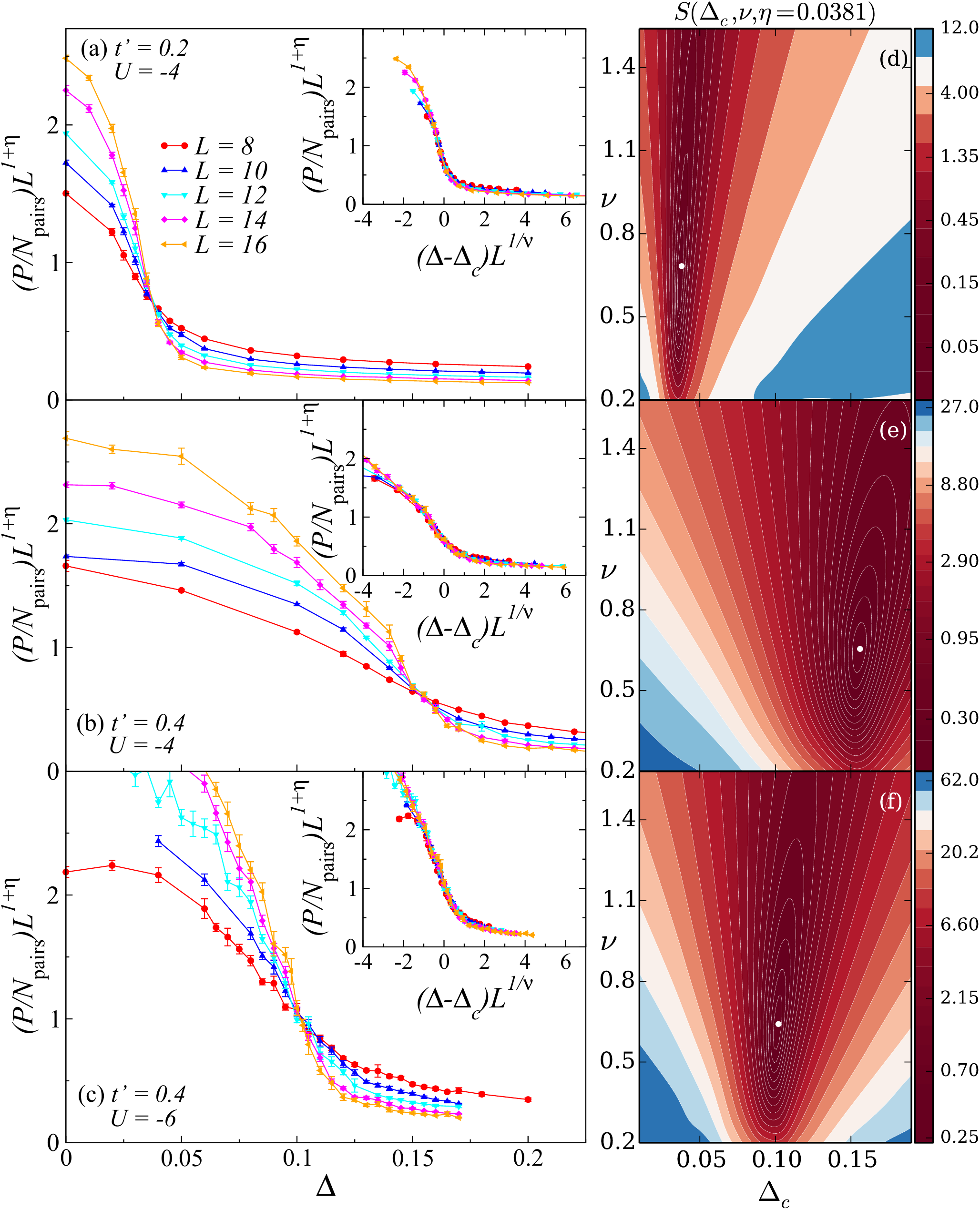}
  \vspace{-0.1cm}
  \caption{(Color online) (a)--(c) show the scaled pair-structure factor as a 
function of $\Delta$ for different values of $U$ and $t'$. The curves 
cross at $\Delta_c$. The insets, show the scaling collapse for the same parameters 
as in the main panels. (d)--(f) show contour plots of the sum of the squared 
residuals $S(\Delta_c,\nu,\eta)$ of fits of $F$ with eight-degree polynomials, 
for the parameters in panels (a)--(c), respectively. We set the value of $\eta$ 
to 0.0381~\cite{Campostrini2006} and found the minima of $S$, as signaled by the 
white dots, for the unknown parameters $\nu$ and $\Delta_c$. $\nu$ at the minima is 
close to the expected $\nu=0.67$ result.}
\label{fig:ps}
\end{figure}

Remarkably, at half-filling, the pair structure factor can be written as
$P=\sum_{i,j}\langle\hat P_i^\dagger\hat P^{}_j\rangle$, i.e., it maps onto 
the zero momentum mode occupation in the hard-core boson model, 
$m_{k=0}=\sum_{i,j}\langle \hat b_i^\dagger \hat b_j^{}\rangle$. In the 
insulating phase of the latter quantum system in 2D one expects 
$\langle \hat b_i^\dagger \hat b_{i+r}^{}\rangle \propto r^{-(1+\eta)} e^{-r/\xi}$
at long distances, as in the corresponding disordered 
3D classical phase, where $\eta$($=0.0381\pm0.0002$~\cite{Campostrini2006}) 
is the anomalous scaling dimension and $\xi$ the correlation length. Near the 
transition, $m_{k=0}$ diverges with $\xi$ as $m_{k=0}\sim\xi^{1-\eta}$ 
\cite{pollet_prokofev_10,Carrasquilla2012} and the fraction $f_0$ of pairs 
that condense in a finite system ($\xi\rightarrow L$) vanishes as 
$f_0\sim L ^{-(1+\eta)}$ \cite{pollet_prokofev_10,Carrasquilla2012}. 
Hence, $f_0$ scales as $f_0L^{1+\eta}=F(|\Delta-\Delta_c|L^{1/\nu})$ 
\cite{Carrasquilla2012}, with $\nu=0.6717\pm0.0001$~\cite{Campostrini2006}. 
Turning back to fermions, we can write
\begin{equation}\label{eq:ansatz}
 (P/N_\text{pairs})L^{1+\eta}=F(|\Delta-\Delta_c|L^{1/\nu}),
\end{equation} 
where the number of pairs is $N_\text{pairs}=L^2/2$.

Figure~\ref{fig:ps} shows the scaled pair-structure factor vs $\Delta$
for two values of $U$ and two values of $t'$. In all cases the curves cross at 
a single point ($\Delta_c$), as expected from the scaling ansatz \eqref{eq:ansatz}. 
That point moves toward larger values of $\Delta$, from Fig.~\ref{fig:ps}(a) to
Fig.~\ref{fig:ps}(b), as $t'$ is increased at constant $U$, and moves toward 
smaller values of $\Delta$, from Fig.~\ref{fig:ps}(b) to Fig.~\ref{fig:ps}(c),
as $|U|$ is increased at constant $t'$. The insets show that, close to the 
crossing points, the data exhibits an almost perfect collapse according to
the scaling ansatz \eqref{eq:ansatz}. To further test this scaling hypothesis, 
we calculate the sum of the squared residuals $S(\Delta_c,\nu,\eta=0.0381)$ of 
fits of $F$ with high-order polynomials (orders 6, 8 and 10) in a fine mesh of 
values for $\Delta_c$ and $\nu$. The value of $\nu$ at which $S$ is minimum
for those polynomials is close to $\nu=0.67$, and the average value is 
$0.65\pm0.08$, $0.66\pm0.04$ and $0.64\pm0.02$~\cite{footnote} for the 
parameters in Figs.~\ref{fig:ps}(a)--\ref{fig:ps}(c), respectively. 
These analyses show that, for those values of $U$, the superconductor 
to Mott-insulator transition in our {\it fermionic} model 
belongs to the 3D $XY$ universality class. This despite the 
fact that the results are for a regime in which $|U|$ is of the order of 1/2 
the bandwidth of the noninteracting system with $\Delta=t'=0$ and, as such, 
strong coupling perturbation theory is not appropriate to describe the system. 
A compilation of crossing points as those in Fig.~\ref{fig:ps} allowed us to 
generate the phase diagram in Fig.~\ref{fig:phase_diagram}(b). Unfortunately, 
for $|U|<4$ and $t'\neq0$, the values of the projection parameters $\Theta$ 
needed to obtain ground-state results are too large and the PDQMC calculations 
become prohibitively long and unstable, so we do not report results in the 
phase diagram for $|U|<4$. Still, for $U=-4$, we can compare the ED and PDQMC 
results. They show that the values of $\Delta_c$ 
in the former are overestimated due to finite size effects.

\subsection{Charge excitations}
\begin{figure}[!tb] 
  \centering
  \includegraphics[width=0.98\columnwidth]{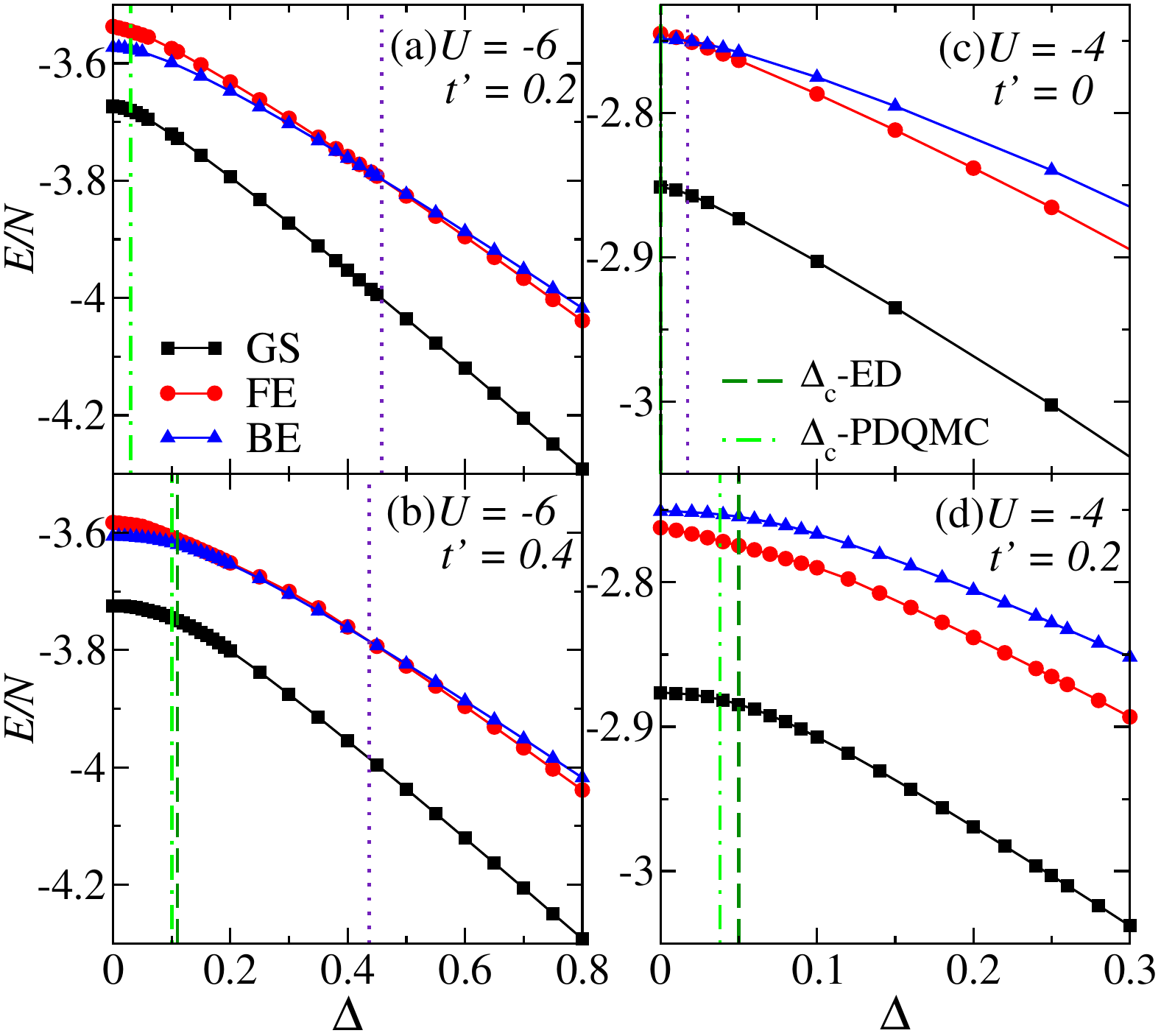}
  \vspace{-0.1cm}
  \caption{(Color online) ED results for the ground state 
  energy (GS), as well as the energies of the first fermionic (FE) and bosonic 
  (BE) charge excitations vs $\Delta$ for different values of $U$ and $t'$. 
  The crossing points between the BE and FE curves (marked by the vertical 
  dotted line) signal a change in the 
  character of the lowest charge excitations. We report the energies per particle, 
  $E/N$ where $N$ is the total number of particles, in $4\times4$ lattices. 
 Dashed and dash-dotted line signal the superconductor-insulator transition 
 using ED and PDQMC, respectively.}
  \label{fig:excitations}
\end{figure}

The observed universality class of the superconductor--Mott-insulator 
transition indicates that the lowest-energy charge excitations have bosonic 
character in {\it both} the superconducting and insulating phases near the 
transition. However, those excitations must be fermionic 
deep in the insulating phase. The insulating region with bosonic low-energy 
excitations is the $s$-wave equivalent of the pseudogap state (a crossover 
regime rather than a thermodynamic phase). In what follows, we explore when the 
lowest charge excitations change from bosonic to fermionic. In 
Fig.~\ref{fig:excitations}, we show the ground-state energy at half-filling 
as well as the energies of the lowest excited states with two extra fermions 
($S^z=0$, lowest energy bosonic excitation) and an extra fermion 
($S^z=1/2$, lowest energy fermionic excitation) as a function of $\Delta$ 
for two values of $U$ and two values of $t'$. 

Figures~\ref{fig:excitations}(a) and \ref{fig:excitations}(b) 
show that, for the values of $\Delta$ at which the superconductor to 
Mott-insulator transition occurs for $U=-6$, the lowest energy excitations 
in both phases are bosonic. However, there is a value of $\Delta>\Delta_c$ 
for which those excitations (within the Mott phase) 
change from bosonic to fermionic. That value of $\Delta$ decreases as $t'$ 
increases [Fig.~\ref{fig:excitations}(a) vs Fig.~\ref{fig:excitations}(b)]
and as $U$ decreases [left vs right panels in Fig.~\ref{fig:excitations}]. 
For $U=-4$ [Figs.~\ref{fig:excitations}(c) and \ref{fig:excitations}(d)],
the ED calculations predict that the transition from bosonic to fermionic 
excitations occurs in the Mott phase for $t'=0$ and in the superconducting
phase for $t'=0.2$. The latter is attributed to finite size effects 
as it contradicts the expectation from the PDQMC results. Also, 
in the weak coupling limit, field-theory arguments anticipate 
the transition to be in the $XY$ universality class 
\cite{nikoli_tesanovic_11,nikoli_11}. These results are nontrivial, 
even though any attractive short-range potential gives rise to bound 
states (Cooper ``molecules'') in 2D, because if interactions are not 
strong enough those ``molecules'' need not be small in comparison to the 
interparticle separation, i.e., bound-state condensation need not occur.

\section{Summary}
We have introduced and studied a 2D model that undergoes a superconductor 
to insulator transition. We determined its phase diagram using ED and PDQMC 
calculations, and showed that, in the parameter regime accessible to PDQMC, 
the transition belongs to the $3D$-$XY$ universality class. Our results echo 
the well-known $XY$ transition of the bosonic Hubbard model, but in a 
{\it fermionic} system. We explored the nature of the lowest energy charge 
excitations and showed that they change from bosonic to fermionic in the 
insulating phase. Our numerical demonstration of ``pseudogap'' physics in a 
realizable model enables a new route for the experimental exploration 
of high-temperature superconductivity using ultracold atoms. While there are 
many microscopic differences between our model and real superconductors, there 
are also several universal similarities that can be exploited. Most notably, 
the dynamics of our system shares a lot in common with charge and vortex dynamics 
near the superconducting transition in layered or quasi-2D superconductors.

\section{Acknowledgments}
This work was supported by the National Science Foundation Grants
No.~PHY13-18303 (R.M.,M.R.) and PHY-1205571 (P.N.), and by CNPq (R.M.). 
The computations were performed in the Institute for CyberScience at Penn State, 
the Center for High-Performance Computing at the University of Southern 
California, and CENAPAD-SP.

\bibliography{attractive_IHMb}

\begin{thebibliography}{44}
\bibitem [{foo()}]{footnote}%
  \BibitemOpen
  \href@noop {} {}\bibinfo {note} {Error bars are defined as the $2\sigma$
  standard deviation of the values of $\nu$, which minimize the sum of square
  residuals for the three polynomials, from the mean.}\BibitemShut {Stop}%
\end{thebibliography}

\end{document}